\documentclass{aastex631}
\usepackage{graphicx}

\shorttitle{EUSO-SPB1 Mission and Science}
\shortauthors{JEM-EUSO Collaboration}
\graphicspath{{./}{figures/}}

\begin{document}

\title{EUSO-SPB1 Mission and Science}

\author{G.~Abdellaoui}
\affiliation{Telecom Laboratory, Faculty of Technology, University Abou Bekr Belkaid, Tlemcen, Algeria} 

\author{S.~Abe}
\affiliation{Nihon University Chiyoda, Tokyo, Japan} 

\author{J.~H.~Adams,~Jr.}
\affiliation{University of Alabama in Huntsville, AL, USA} 

\author{D.~Allard}
\affiliation{Universit\'e de Paris, CNRS, AstroParticule et Cosmologie, Paris, France} 

\author{G.~Alonso}
\affiliation{Universidad Polit\'ecnia de Madrid (UPM), Madrid, Spain} 

\author{L.~Anchordoqui}
\affiliation{Lehman College, City University of New York (CUNY), NY, USA}

\author{A.~Anzalone}
\affiliation{INAF - Istituto di Astrofisica Spaziale e Fisica Cosmica di Palermo, Italy}
\affiliation{Istituto Nazionale di Fisica Nucleare, Sezione di Catania, Italy} 

\author{E.~Arnone}
\affiliation{Dipartimento di Fisica, Universita' di Torino, Italy}
\affiliation{Istituto Nazionale di Fisica Nucleare, Sezione di Torino, Italy}

\author{K.~Asano}
\affiliation{Institute for Cosmic Ray Research, University of Tokyo, Kashiwa, Japan}

\author{R.~Attallah}
\affiliation{LPR at Department of Physics, Faculty of Sciences, University Badji Mokhtar, Annaba, Algeria}

% ...

\collaboration{0}{JEM-EUSO}

\correspondingauthor{L.~Wiencke}
\email{lwiencke@mines.edu}

\begin{abstract}
The Extreme Universe Space Observatory on a Super Pressure Balloon 1 (EUSO-SPB1) was launched in 2017 April from Wanaka, New Zealand. The plan of this mission of opportunity on a NASA super pressure balloon test flight was to circle the southern hemisphere. The primary scientific goal was to make the first observations of ultra-high-energy cosmic-ray extensive air showers (EASs) by looking down on the atmosphere with an ultraviolet (UV) fluorescence telescope from suborbital altitude (33~km). After 12~days and 4~hours aloft, the flight was terminated prematurely in the Pacific Ocean. Before the flight, the instrument was tested extensively in the West Desert of Utah, USA, with UV point sources and lasers. The test results indicated that the instrument had sensitivity to EASs of ${\gtrapprox 3}$~EeV. Simulations of the telescope system, telescope on time, and realized flight trajectory predicted an observation of about 1 event assuming clear sky conditions. The effects of high clouds were estimated to reduce this value by approximately a factor of 2. A manual search and a machine-learning-based search did not find any EAS signals in these data. Here we review the EUSO-SPB1 instrument and flight and the EAS search. 
\end{abstract}

\section{Introduction}\label{sec:intro}
The sources and acceleration mechanisms that produce the highest energy particles ever observed remain unknown.
With measured energies that can exceed 100~EeV~\citep{Abu_Zayyad_2013, 2018AJ....156..123A}, ultra-high-energy cosmic rays 
(UHECRs\footnote{By convention, UHECRs are 
cosmic-rays with energies above 1 EeV ($10^{18}$~eV).}) 
occupy a tantalizing position in the multi-messenger view of the cosmos.
The 100~EeV scale is six orders of magnitude above the highest energy gamma rays~\citep{Cao2021} and five orders of magnitude above the highest energy neutrinos~\citep{Aartsen2021, PhysRevD.104.022001} observed to date. 
Because UHECRs are charged, they are the only known high energy multi-messengers that can be accelerated directly by the sources of interest. 
The low flux of UHECRs requires indirect measurement techniques that use the atmosphere as a giant calorimeter. UHECRs that reach Earth's atmosphere convert their kinetic energy into extensive air showers (EASs), which generate fluorescence light via excitation of atmospheric nitrogen. The atmospheric fluorescence technique has been developed and used successfully by Fly's Eye~\citep{1994ApJ...424..491B}, the High-Resolution Fly's Eye (HiRes)~\citep{PhysRevLett.100.101101}, the Pierre Auger Observatory~\citep{2015172}, and Telescope Array~\citep{ABBASI2016131} to measure EASs in 3D from the ground. The atmosphere is, by definition, the largest-volume calorimeter on Earth.

The Extreme Universe Space Observatory on a Super Pressure Balloon 1 (EUSO-SPB1) was the Joint Experiment Missions
for Extreme Universe Space Observatory (JEM-EUSO) collaboration's first mission targeting UHECR EASs by looking down on the atmosphere from suborbital space. This mission represents an important step toward establishing a UHECR detector in space~\citep{Bensen1981,STECKER2004433,Krizmanic2013,Haungs2015,Bertaina:2019+0} that would view a much larger (${\approx}$100${\times}$) atmospheric footprint, map the entire sky at extreme energies, and discover the sources of UHECRs. The Probe of Extreme Multi-Messenger Astrophysics (POEMMA) ~\citep{Olinto_2021} would also have sensitivity to ultra-high-energy photons, monopoles and super heavy forms of dark matter, and, via interactions in Earth's limb, very high energy neutrinos~\citep{PhysRevD.101.023012}. 
\begin{figure}
\centering
\includegraphics[width=6.5in,keepaspectratio]{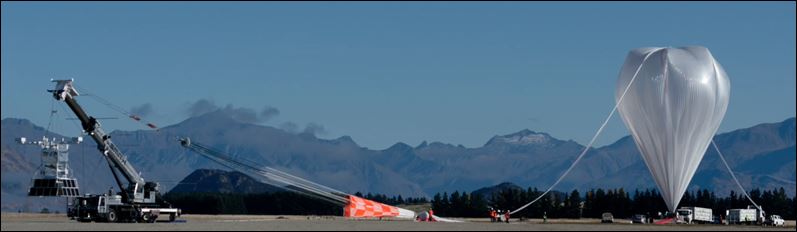} 
 \caption{EUSO-SPB1 shortly before its 2017 launch from Wanaka, New Zealand. The instrument hangs below the crane on the far left.}
\label{fig:launch}
\end{figure}

EUSO-SPB1 flew as a NASA mission of opportunity. It was suspended below a NASA super pressure balloon that was launched as a test flight on 2017 April 24 23:51~UT from Wanaka, New Zealand (Figure~\ref{fig:launch}). The EUSO-SPB1 science goals included the following:
\begin{enumerate}
\itemsep0em
\item making the first observations of UHECR EASs by looking down from suborbital space with an air fluorescence detector,
\item measuring background UV light at night over ocean and clouds, and
\item searching for fast UV pulse-like signatures from other objects.
\end{enumerate}

\section{Ultra-Long-Duration Scientific Balloon Flights from Wanaka, New Zealand} \label{sec:ULDBNZ}

\begin{figure}
\centering
\includegraphics[width=6.0in,keepaspectratio]{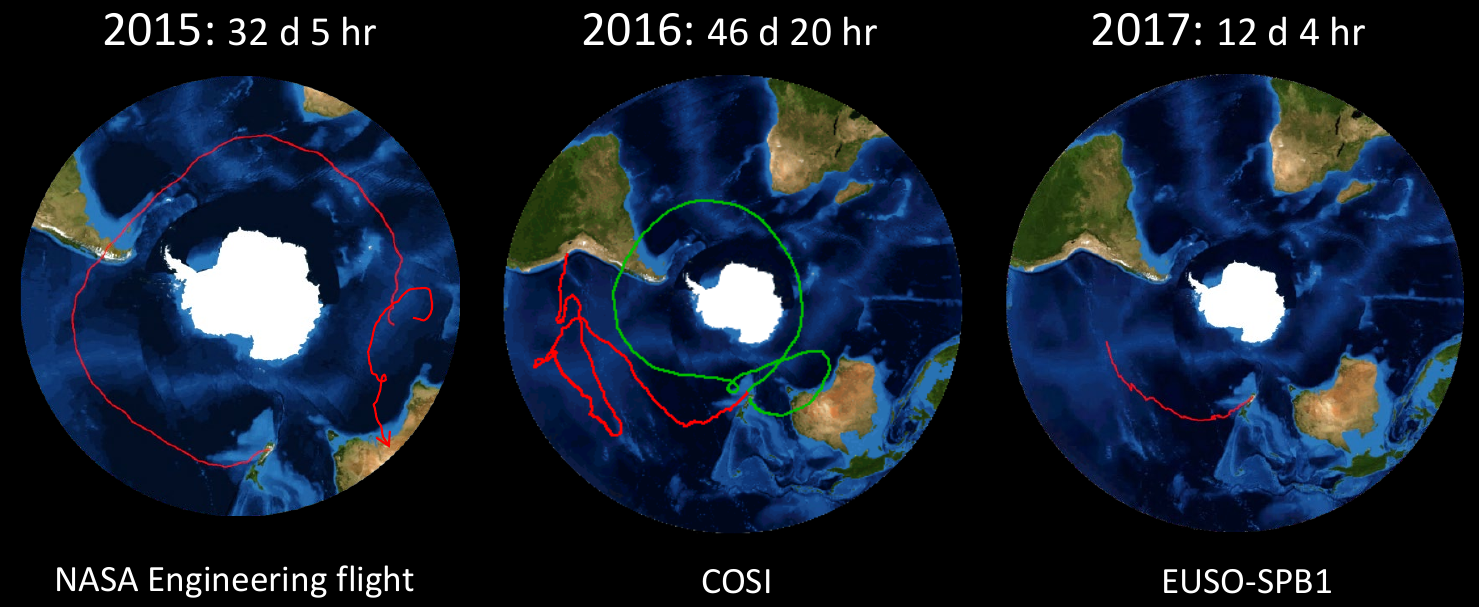} 
\caption{Trajectories of the NASA super pressure balloon test flights launched from Wanaka, New Zealand through 2017. The green portion of a trajectory denotes a completed circumnavigation}
\label{fig:trajectories}
\end{figure}

Located at 45$^{\circ}$S latitude, Wanaka lies below a fast stratospheric air circulation that develops twice a year
about 33~km (7~mbar) above the Southern Ocean. This circulation flows easterly in the southern autumn and westerly in the southern spring.
Super pressure balloons (SPBs) 
are designed to float at a constant displacement volume and consequently at a constant altitude, even at night. Thus a payload launched from Wanaka under a specially prepared SPB could ride this stratospheric circulation for months, completing multiple suborbital circumnavigations followed by controlled termination over land. Like conventional zero-pressure stratospheric balloons, SPBs also drift with the balloon-altitude wind currents. NASA's first engineering SPB mission launched from Wanaka~\citep{cathey2017performance} flew for 32~days in 2015 and landed in Australia. The first scientific payload, the Compton Spectrometer and Imager (COSI)~\citep{Lowell_2017}, flew the following year for 46~days, circled the Southern Ocean, and landed in Peru. The 2017~flight was the third SPB test flight from Wanaka, with a nominal duration target of 100 days. The paths of these three~flights are shown in Figure~\ref{fig:trajectories}.

\begin{figure}
\includegraphics[width=6.5in,keepaspectratio]{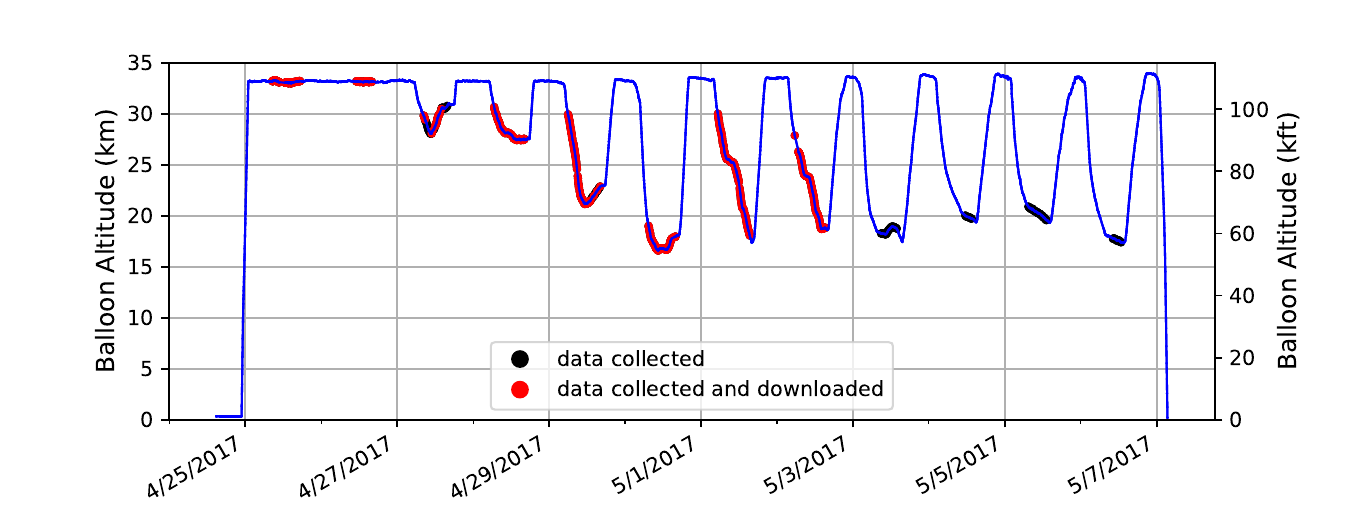}
\caption{Altitude profile of the 2017~mission as a function of UT date. The instrument was operated at night when the moon was below the horizon. The altitude fluctuations that started after 2 days aloft were caused by a leak that prevented the balloon from maintaining a fully inflated super-pressure state at night.}
\label{fig:traj_height}
\end{figure}

Unfortunately, the 2017~balloon developed a helium leak that necessitated an early controlled termination of the mission (Figure~\ref{fig:traj_height}). The entire flight train was ``valved down" about 300~km southeast of Easter Island 12~days after launch. It currently rests on the deep ocean floor.

\begin{figure}[ht]
\centering
\includegraphics[width=6.0in,keepaspectratio]{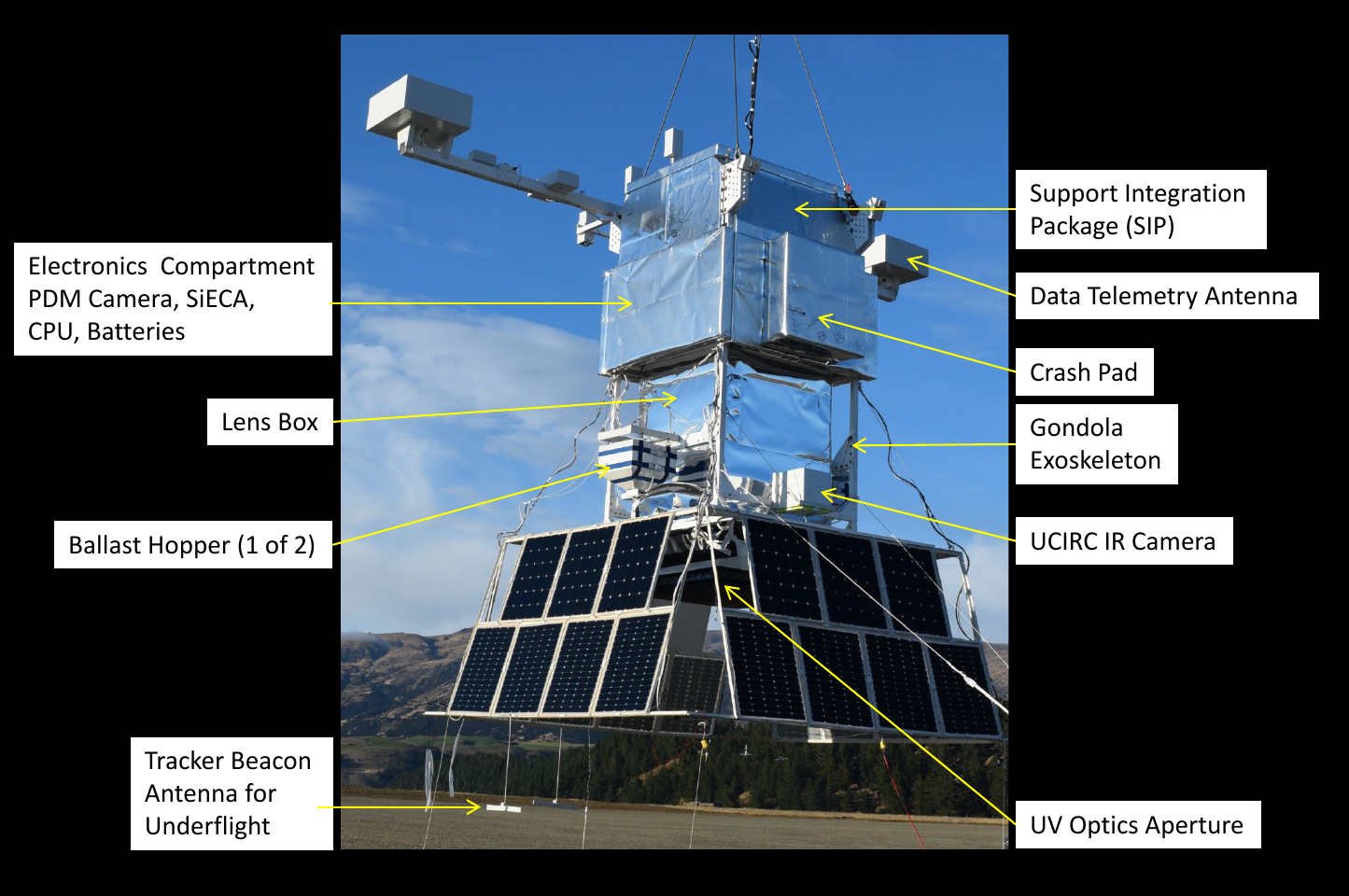} 
 \caption{EUSO-SPB1 instrument in flight-ready configuration. See text for description of subsystems.}
 \label{fig:instrument}
\end{figure}

Despite the setbacks, the EUSO-SPB1 instrument (Figure~\ref{fig:instrument}) operated successfully while aloft and returned about 60~GB of data. Here we describe the instrument, preflight testing in the laboratory and desert, the mission, the data, and the search for EASs. Preparations for the EUSO-SPB2 mission (Section~\ref{sec:Future Missions}) are underway.

\section{EUSO-SPB1 Instrument}
\subsection{2014 instrument and mission}
The 2017~EUSO-SPB1 mission succeeded the 2014 EUSO-Balloon overnight mission sponsored by the French Space Agency and launched from Timmins, Ontario, Canada, by the Canadian Space Agency on a zero-pressure balloon. The fluorescence telescope (FT) camera was triggered externally by an onboard 20~Hz clock. The instrument recorded UV terrestrial emission levels~\citep{ABDELLAOUI201954} and sampled UV flashes and UV laser tracks generated by light sources flown below the balloon on a helicopter. This mission yielded the first observation of UV tracks by a fluorescence telescope looking down on the atmosphere~\citep{Abdellaoui_2018}. The entire flight train splashed down in a small lake and the instrument was recovered intact.

\subsection{2017~instrument}
\begin{figure} [t]
\centering
\includegraphics[width=6.0in,keepaspectratio]{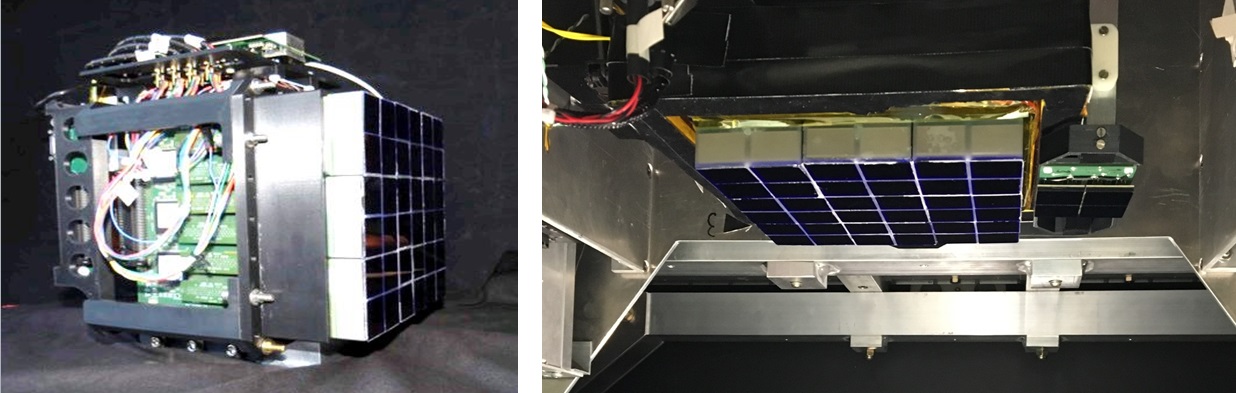} 
 \caption{Left: The photodetector module (PDM) of the EUSO-SPB1 instrument. Right: as installed at the focal plane. The SiECA prototype is positioned to the right of the PDM.}
 \label{fig:focal_surface}
\end{figure}

The 2014 instrument was upgraded extensively for the 2017 launch. Specifications of the 2017~instrument~\citep{Bacholle:2017JL} are listed in Table~\ref{tab:spec}. The upgrades featured a trigger to identify EAS candidates, a new set of Fresnel lenses, a new UV camera (Figure~\ref{fig:focal_surface}) with higher quantum efficiency multi-anode photomultiplier tubes (MAPMTs), an integrated high voltage (HV) system, a new flight CPU, interfaces to the NASA Support Integration Package (SIP) and telemetry, new control software, thermal sensors interfaced to the SIP, a solar power system, and a gondola exoskeleton frame on which all equipment was mounted, including an antenna boom. The solar power system was designed to support the mission through the southern winter including a trajectory excursion to about 60\arcdeg S. 

\begin{table}[htp]
\centering
\begin{tabular}{| l l l l|}
\hline
Item & Specification & Notes&\\
\hline
Energy threshold & ${\approx}$3~EeV & 50\%\ trigger threshold & \\
Trigger aperture & ${\approx}$20~km$^{2}$sr (5~EeV) & &\\
 &${\approx}$200~km$^{2}$sr (10~EeV) & At 33 km altitude & \\
Telescope optics & 2${\times}$1 m$^2$ Fresnel lenses& PMMA &\\
Field of view & 11.1\arcdeg${\times}$11.1\arcdeg & From stars, lasers&\\
Pixel field of view & 0.2\arcdeg${\times}$0.2\arcdeg& For active area&\\
Pixel ground footprint & 120~m${\times}$120~m & As projected from 33 km&\\
Number of pixels & 2304 (48${\times}$48)& 36~MAPMTs${\times}$64 pixels each&\\
MAPMT & R11265-113-M64-MOD2~& Hamamatsu &\\
UV transmitting filter & BG-3,~2~mm thick& 1 per MAPMT &\\
Readout & DC coupled & 100~MHz double-pulse resolution &\\
Time-bin duration & 2.5~${\mu}$s integration & Event packet = 128~bins (320~${\mu}$s)&\\
\hline
Balloon &18${\times}10^{6}$ ft$^{3}$~ (0.5${\times}10^{6}$ m$^{3}$)&Helium & \\
Nominal float height & 33.5 km (110000~ft) & &\\
Telemetry (data) & 2${\times}{\approx}$75~kbits~${s^{-1}}$& 2~Iridium OpenPort& \\
Telemetry (comms) & ${\approx}$1.2~kbits~${s^{-1}}$ (255~bit bursts)&2~Iridium Pilots& \\
Power consumption & 40~W (day), 70~W (night)& Includes 20~W heater&\\
Batteries &10, each 42 A${\cdot}$h &Odyssey PC1200 12 V lead acid&\\
Solar panels & 3${\times}$100~W on all 4 sides & SunCat Solar &\\
Detector weight & 1223~kg (2250~lb) &Without SIP, antennas, and ballast & \\
Releasable ballast & 545~kg (1200~lb)& ${\le}$50~lb remaining at termination & \\
Total weight & 2500~kg (5500~lb)&Everything below balloon&\\
\hline
Flight start &2017 April 24 23:51~UT ~& 44.7218\arcdeg S 169.2540\arcdeg E&\\
Flight end &2017 May 6 3:40~UT ~& 29.3778\arcdeg S 106.5037\arcdeg W& \\
Flight duration & 12~days 4~hr & & \\

\hline
\end{tabular}
\caption{Specifications of EUSO-SPB1 and 2017~mission.}
\label{tab:spec}
\end{table}

\begin{figure} [ht] 
\centering 
\includegraphics[width=4.5in,keepaspectratio]{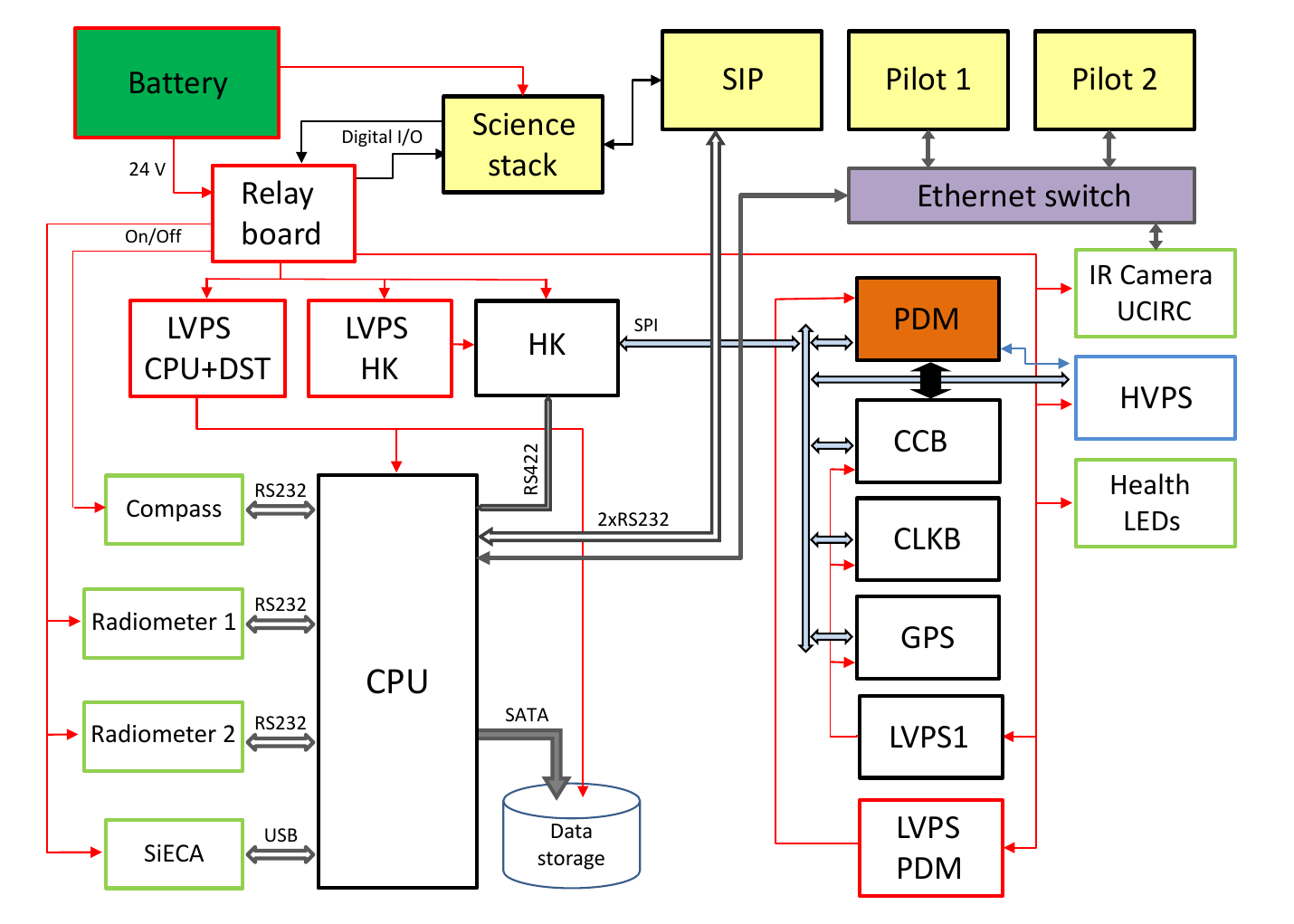}
\caption{EUSO-SPB1 data readout system and its connections to the various subsystems.}
\label{fig:dp_diagram}
\end{figure}

 EUSO-SPB1 was designed to operate at stratospheric altitude during nights with little or no moon to detect UV tracks from EASs. The subsystem architecture is diagrammed in Figure~\ref{fig:dp_diagram}. Two 1 m${^2}$ polymethyl methacrylate (PMMA) Fresnel lenses focus light from below onto a UV-sensitive, custom high-speed camera. The focal surface of the EUSO-SPB1 telescope features a photodetector module (PDM) that counts single photoelectrons (SPEs). For assembly purposes, 4~MAPMTs, of 64~channels each, are covered by a square BG-3~UV-transmitting optical filter to form an elementary cell (EC). Nine ECs in a 3${\times}$3 arrangement make up one PDM. Located in the PDM behind the ECs are six circuit boards that count the numbers of photoelectrons. These boards plug into a central control buffer board that hosts the VHDL trigger logic.

 The PDM is operated with ${-}$1100~V applied to the photocathodes. The nominal gain of the MAPMTs is 10$^{6}$. A Cockcroft-Walton circuit~\citep{Plebaniak:2017XU} generates the HV. This circuit is implemented in a board potted into each EC subassembly. The MAPMT anodes are held at ground and coupled directly to the digitization electronics. This DC coupling permits photometric calibrations using pulsed or DC light sources and also permits direct measurements of background light levels. The digitization electronics identify the small current pulses that are initiated by the emission of single photoelectrons from the photocathodes. The double pulse resolution of the digitization electronics is about 6~ns. The number of SPE counts in each MAPMT channel is tallied in 2.5~${\mu}$s time bins. One time bin is referred to as a gate time unit (GTU) in this paper.

The onboard trigger system~\citep{ABDELLAOUI2017150, Bayer:2017J2} operates at the MAPMT level and scans buffered pixel count lists for locally persistent signals above background as averaged over a specified time, within 3${\times}$3 pixel cells. Persistence settings of 1, 2, and 5 GTUs were used during the flight. A persistence setting of 2 GTUs, for example, corresponds to 5~${\mu}$s. The local background threshold level is adjusted dynamically to significantly reduce the number of fake triggers caused by slowly moving objects such as airplanes, and electrical storms. On receipt of a trigger, the data processing (DP) system~\citep{SCOTTI201994} copies 128~consecutive data frames (40 before and 88 after the trigger for a total of 320~$\mu$s) from a system buffer to an onboard 1~TB raid array. A single data frame contains a list of the number of photoelectrons recorded in each of the 2304 individual pixels over the same 2.5~$\mu$s interval. The DP system is also interfaced to ancillary systems, as diagrammed in Figure \ref{fig:dp_diagram}. The flight computer hosts a comprehensive modular control software package~\citep{Fornaro:2019} that can be readily configured for testing and flight. Field tests of the trigger system are described in Section~\ref{sec:FieldTests}.

\begin{figure}
\centering
\includegraphics[width=5.0in,keepaspectratio]{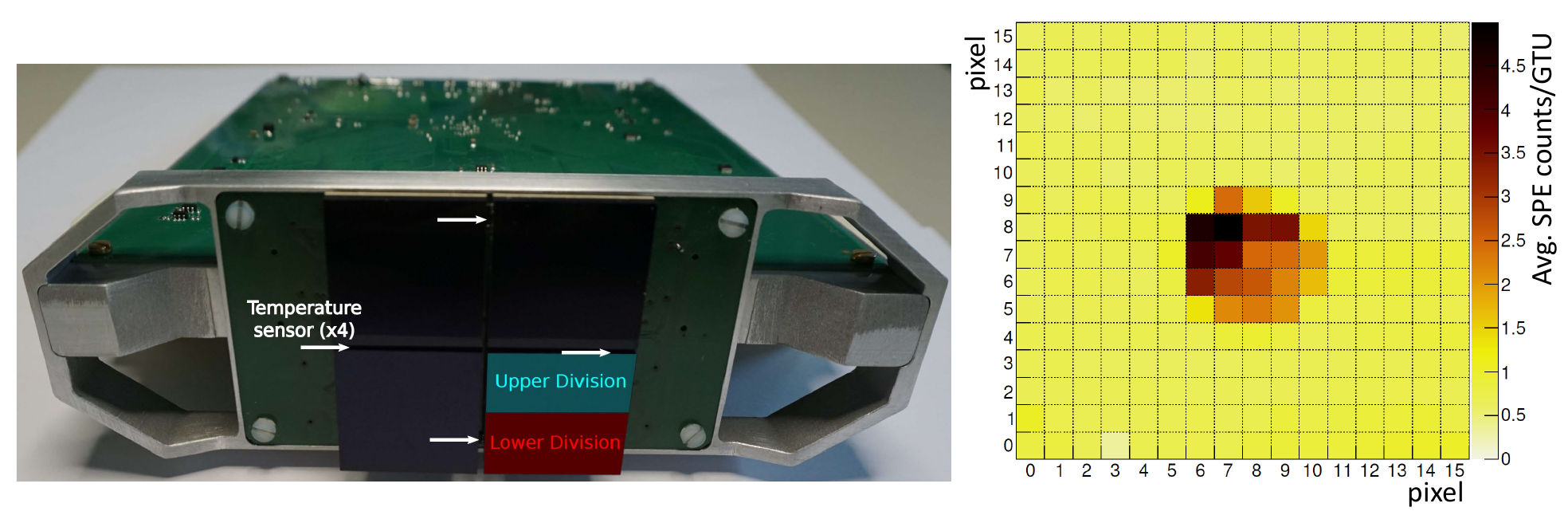}
\caption{Left: SiECA 256-channel SiPM array of four~64-channel Hamamatsu 
S13361-3050AS-08 sensors was flown as a test on the EUSO-SPB1 mission. Right: laboratory data recorded from an uncollimated DC light spot source.}
\label{fig:sieca}
\end{figure}

As a technology test of silicon photomultipliers (SiPMs) in near space, the EUSO-SPB1 focal surface also includes a 256-channel silicon photomultiplier elementary cell array (SiECA)~\citep{Painter:2017Vw, Renschler2018CharacterizationOH} mounted next to the MAPMTs of the PDM (Figures~\ref{fig:focal_surface} and~\ref{fig:sieca}). This add-on system was flown in a stand-alone sampling mode.

An infrared (IR) camera system was developed and flown to record IR images of the scene below the balloon to identify clouds and estimate the heights of the cloud tops. Measuring clouds is important because high clouds reduce the instantaneous aperture to detect EASs by the EUSO-SPB1 instrument~\citep{Adams2015}. The University of Chicago Infrared Camera (UCIRC)~\citep{Rezazadeh:2017RJ} featured two identical IR cameras that pointed down toward the same region. The field of view was 24${^\circ}\times$30${^\circ}$. Each camera had an IR filter. One filter transmitted IR light between wavelengths of 11.5 and 12.9~$\mu$m. The other filter transmitted IR light between wavelengths of 9.6 and 11.6~$\mu$m. These values were selected because they fall near the typical blackbody peak for clouds. The method for measuring cloud color temperature from which the cloud top height can be derived is described elsewhere~\citep{Anzalone:2019}.

The mechanical upgrades to the payload complied with NASA requirements for balloon gondolas, with the added requirement that the overall height, including antennas, be lower than the door of the aircraft hangar from which the mission is staged. The NASA SIP and antennas were mounted on top of the gondola frame, and two ballast hoppers were mounted on opposite sides of the frame to allow the UV fluorescence telescope and IR camera system an unobstructed downward field of view. The fluorescence telescope module could be rolled into the gondola structure and attached in about 30~minutes. A four-sided solar array and a light baffle were connected to the payload outside the hangar.

\section{Desert Field Tests}
\label{sec:FieldTests}

\begin{figure}
\centering
\includegraphics[height=3.5in,keepaspectratio]{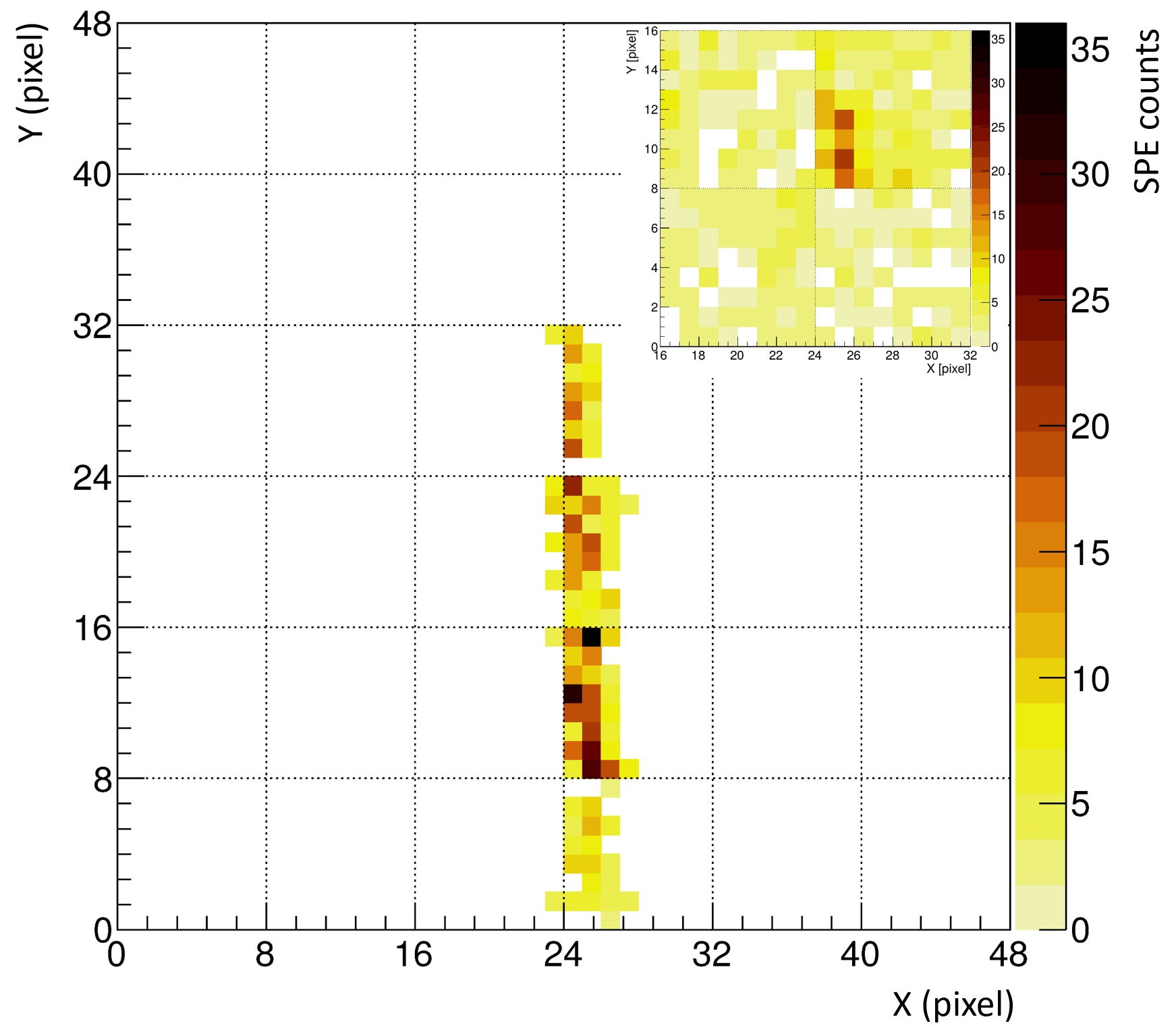} 
\caption{Example of a 1.3 mJ 355 nm laser track recorded during desert field tests of EUSO-SPB1. The laser steering optics were aimed at an elevation angle 45\arcdeg~away from the telescope position. The top of the track corresponds to a height of 9.3 km and viewing distance of 35 km. The pixels displayed are a sum over ten 2.5~${\mu}$s frames of the trigger signal. A single 2.5~${\mu}$s frame that captures the moving light spot from the laser is shown in the inset.}
\label{fig:laser_track_gtus}
\end{figure}

\begin{figure}
\centering
\includegraphics[height=2.5in,keepaspectratio]{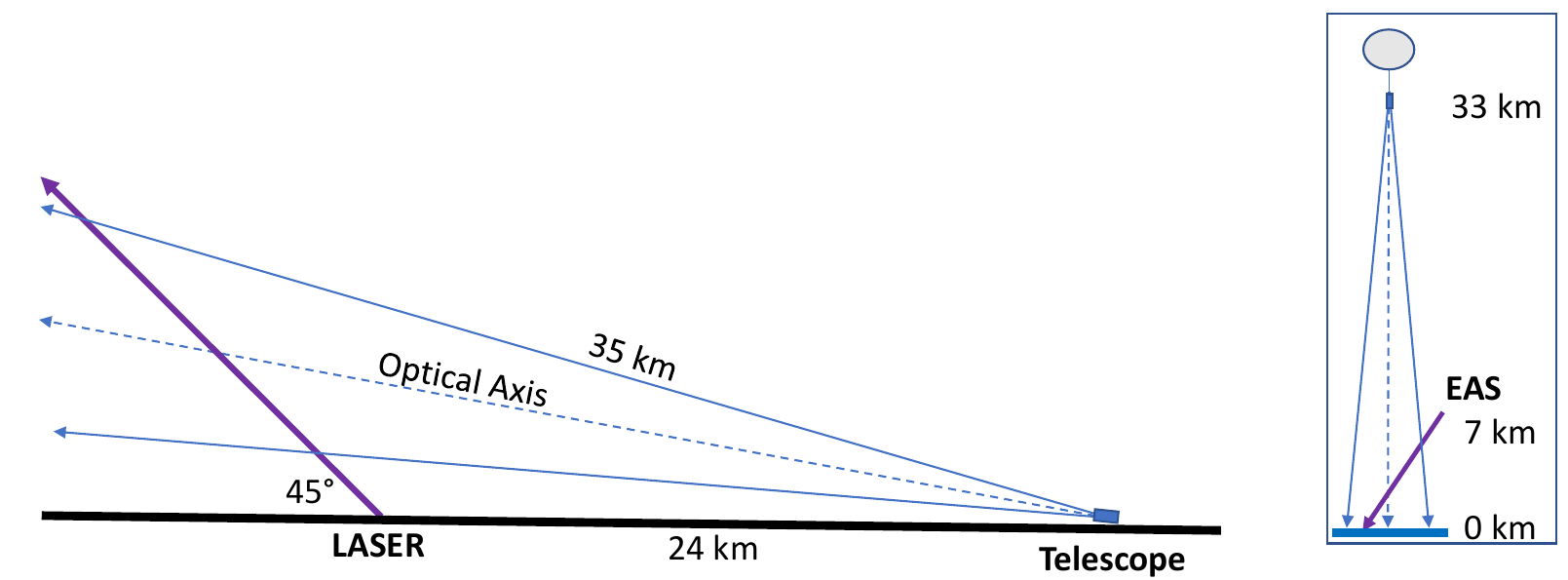} 
\caption{Arrangement the EUSO-SPB1 fluorescence telescope and a laser during field tests is shown on the left. The diagram on the right shows this arrangement after rotation by 90\arcdeg, so that the telescope optical axis is pointing down, as it did under the balloon. The laser and EAS axes are in the same geometrical position relative to the telescope optical axis. Both diagrams are side views.}
\label{fig:laser_diagram}
\end{figure}

\begin{figure}
\centering
\includegraphics[height=3.0in,keepaspectratio]{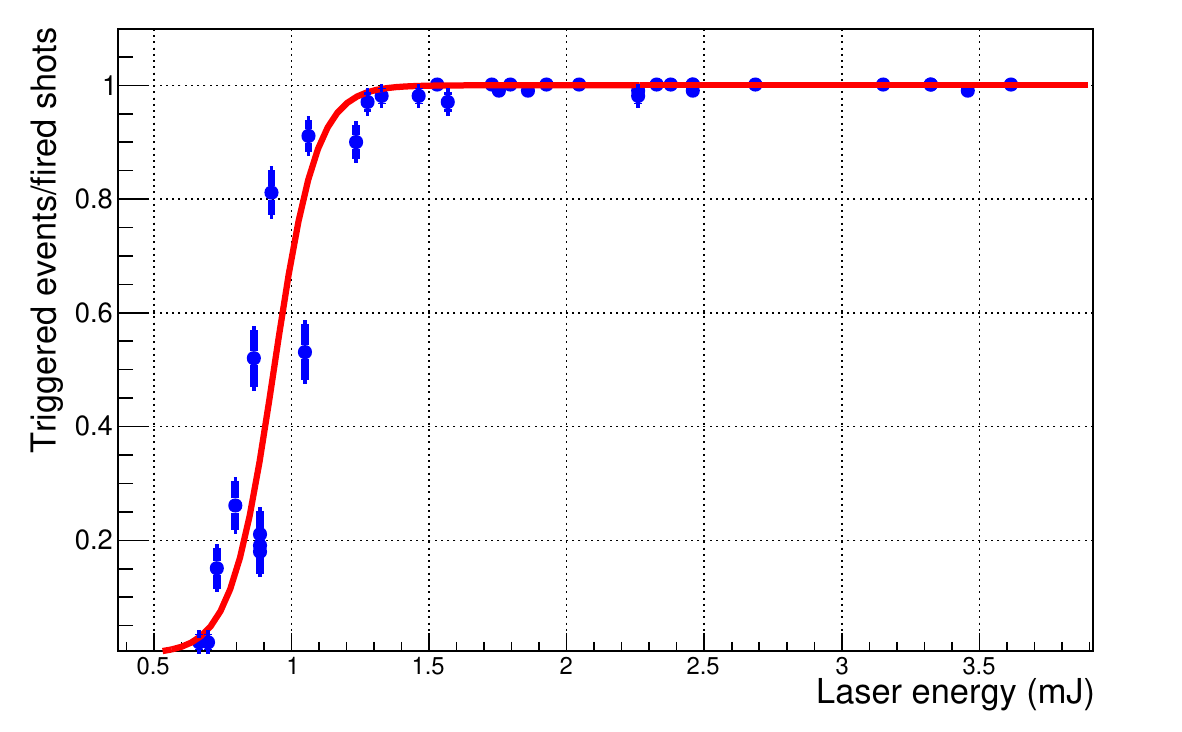} 
\caption{Trigger efficiency of the fluorescence telescope vs. laser energy as measured by sweeping the energy of a distant UV laser. A total of 100~shots were fired at each energy setting. See text for details.}
\label{fig:laser_energy_scan}
\end{figure}

The EUSO-SPB1 fluorescence telescope system was tested in the laboratory and in the West Desert of Utah, USA~\citep{Adams2021}, at the Telescope Array site. Measurements of a 365 nm calibrated point source on a mast yielded an estimate of end-to-end absolute photometric calibration as 0.10${\pm}$0.01 SPE counts per incident photon. This value is comparable to a laboratory piecewise calibration. A pulsed UV laser system~\citep{Hunt:2016do} having a 10 ns pulse width was placed 24~km from EUSO-SPB1 and used to measure the trigger efficiency to speed-of-light tracks in the atmosphere. Like an EAS, the pulsed laser produces a moving spot of UV light traveling at the speed of light. Laser measurements with the beam tilted 45$^{\circ}$ away from the telescope position were recorded (Figure \ref{fig:laser_track_gtus}) to approximate geometrically the expected distance from EUSO-SPB1 at float altitude to an EAS of a typical inclination traversing the telescope field of view below. A comparison of these configurations is diagrammed in Figure \ref{fig:laser_diagram}. The geometrical equivalence means that the rate of travel of the light spot crossing the camera is equivalent in the two orientations, as is the 1/r effect for a line source where r is the distance between
the telescope and the laser pulse. The trigger efficiency was found to be 50\% for a laser energy of 0.94${\pm}$0.02~mJ and approached 100\% efficiency at 1.5~mJ (Figure \ref{fig:laser_energy_scan}). The data were collected in two energy sweeps over about 3 hours. A change in the atmospheric clarity during this period across the 24 km separating the laser and detector is the most likely reason that the points in the threshold range are separated beyond the error bars which are statistical. The data points to the left and right of the fitted curve in the region below 50\% trigger efficiency are separated by about two hours in time.

The inclined 0.8 mJ laser track, as viewed from the side in the desert field test, was estimated to appear about as bright as an inclined EAS of 3 EeV would appear when the fluorescence telescope was looking down from the balloon float altitude. This estimate assumed a nominal aerosol optical depth for the desert laser measurement. 

Noise trigger rates were typically at the level of 1~Hz over the full camera during the field tests. These trigger rates increased slightly when bright stars or meteors crossed the field of view, or when there was lightning outside the field of view.

\section{EUSO-SPB1 Campaign and Instrument Monitoring}

After the EUSO-SPB1 components reached New Zealand in late January 2017, the payload was reassembled. To obtain a final flat-field calibration, the payload was suspended from a crane above an illuminated Tyvek screen at night and rotated in azimuth while the FT collected a data set of full-camera triggers. The instrument was declared flight-ready on 2017 March 25. Following seven aborted attempts, the payload was launched successfully on April 24 23:30~UT, 1 day before the new moon. On reaching the New Zealand coast, the balloon drifted northward, passing about 30 km east of Christchurch before heading out over the Pacific Ocean. Instrument monitoring and operations were handled through centers in Europe, Japan, and the USA. To facilitate downloading, telescope data files were limited to a duration of 2 minutes. A shorter (30 s) file was recorded at the start of each hour and downloaded with highest priority to provide telescope monitoring information.

\begin{figure} 
\centering
\includegraphics[width=5.5in,keepaspectratio]{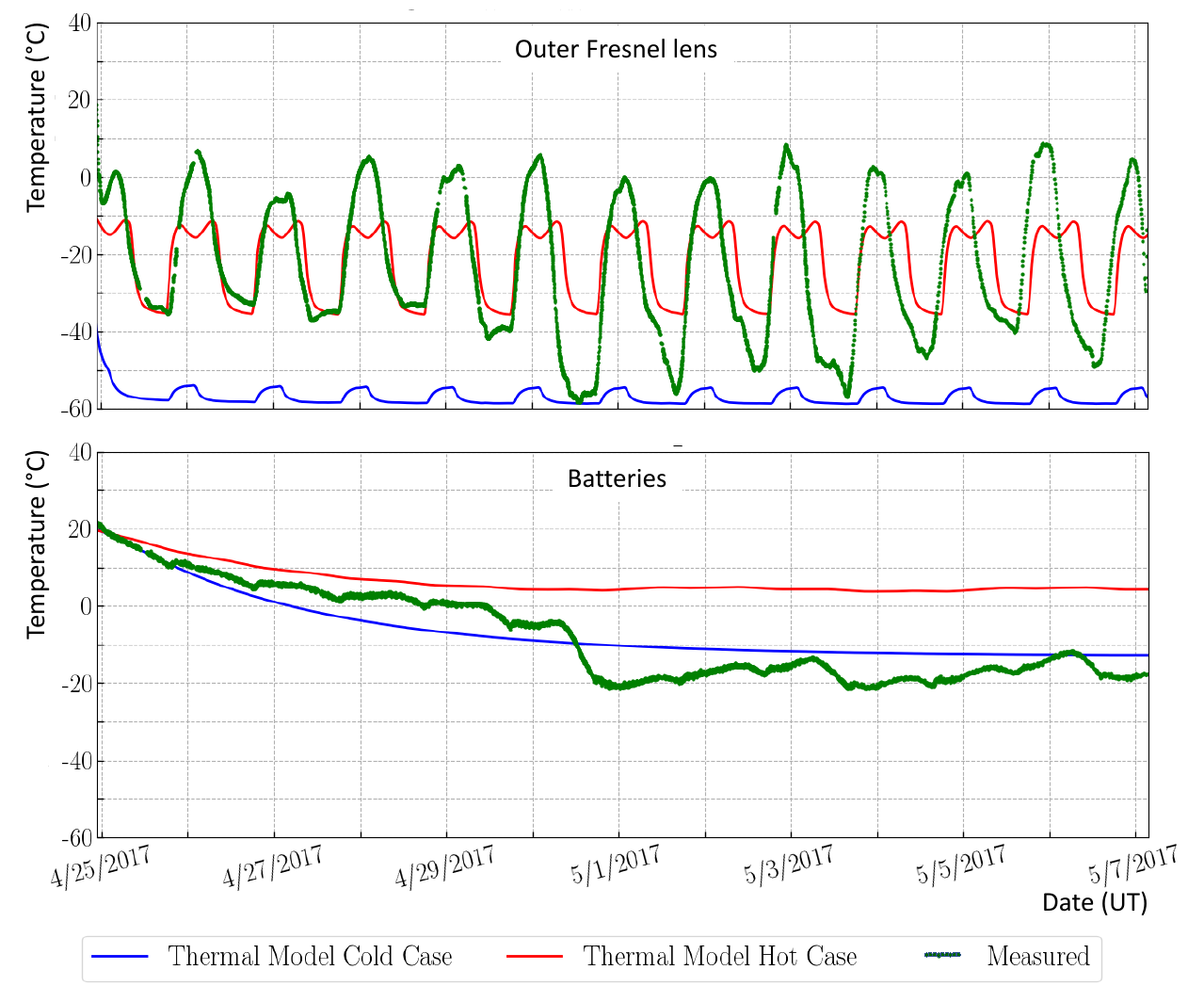} 
\caption{Examples of temperature data (top panel: outside the Fresnel lens at the telescope optics aperture; bottom panel: batteries) recorded during the flight compared with predictions of the preflight thermal model (see text).}
\label{fig:Thermal}
\end{figure}

 Examples of thermal monitoring measurements recorded during the flight are compared with predictions of the instrument preflight thermal model in Figure \ref{fig:Thermal}. This model was developed to predict the hottest and coldest cases for a long-duration flight at 33~km altitude. The hot case assumed a 45$^{\circ}$S latitude flight and a March~1 launch, whereas the cold case assumed a 65$^{\circ}$S latitude with a June~22 launch. The largest excursions of the data below the warm case prediction started on April~30, when the balloon did not maintain a super-pressure state at night and the payload descended at night to 18~km, reaching the colder air of the tropopause. The excursions of the front lens temperature above the warm case may be due the model underestimating the heat transfer effect of direct sunlight on the telescope walls and/or from indirect sunlight reflecting from the ocean and clouds onto the front lens.

\begin{figure} 
\centering
\includegraphics[width=6.0in,keepaspectratio]{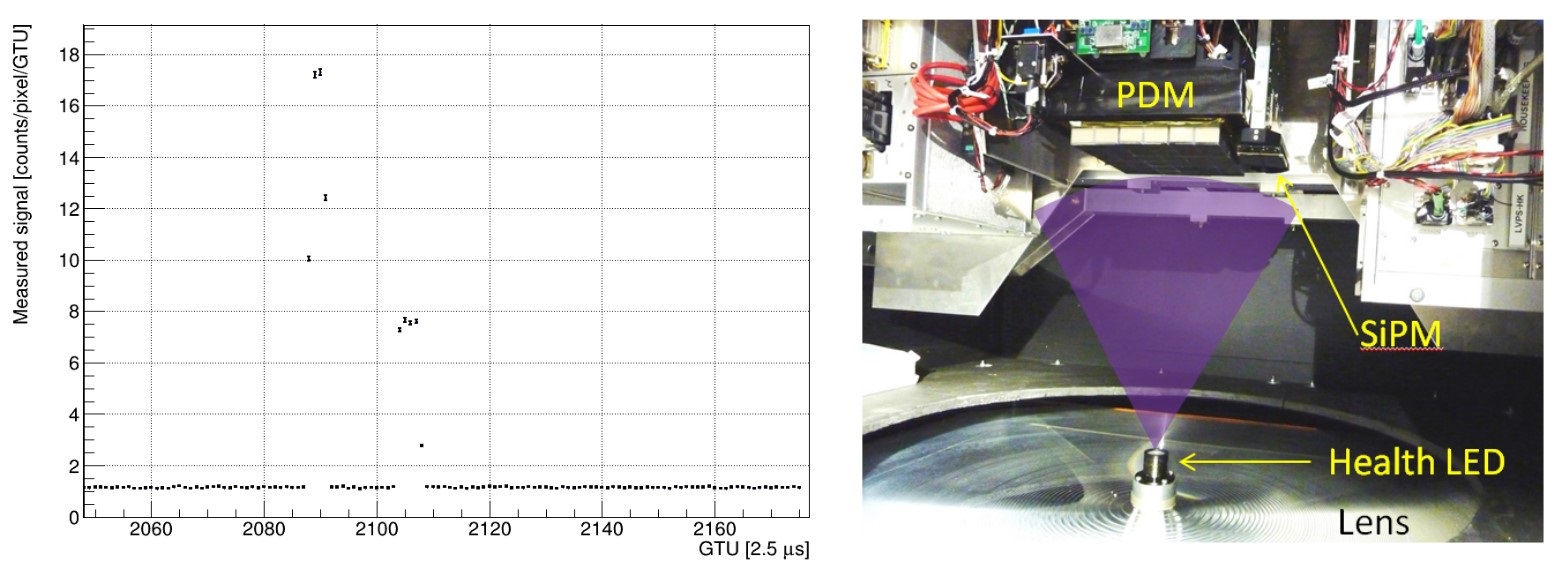}
\caption{Left: A double light pulse generated by the onboard health LED system as recorded by EUSO-SPB1. The response shown here gives the SPE counts/pixel averaged over all pixels as a function of (2.5~$\mu$s) GTU number. The error bars represent the corresponding standard deviation about each average divided by the square root of the number of pixels in the camera.  Right: The health LED was mounted in the middle of the Fresnel lens in front of the UV camera. }
\label{fig:Health_LED}
\end{figure}

\begin{figure} 
\centering
\includegraphics[width=6.5in,keepaspectratio]{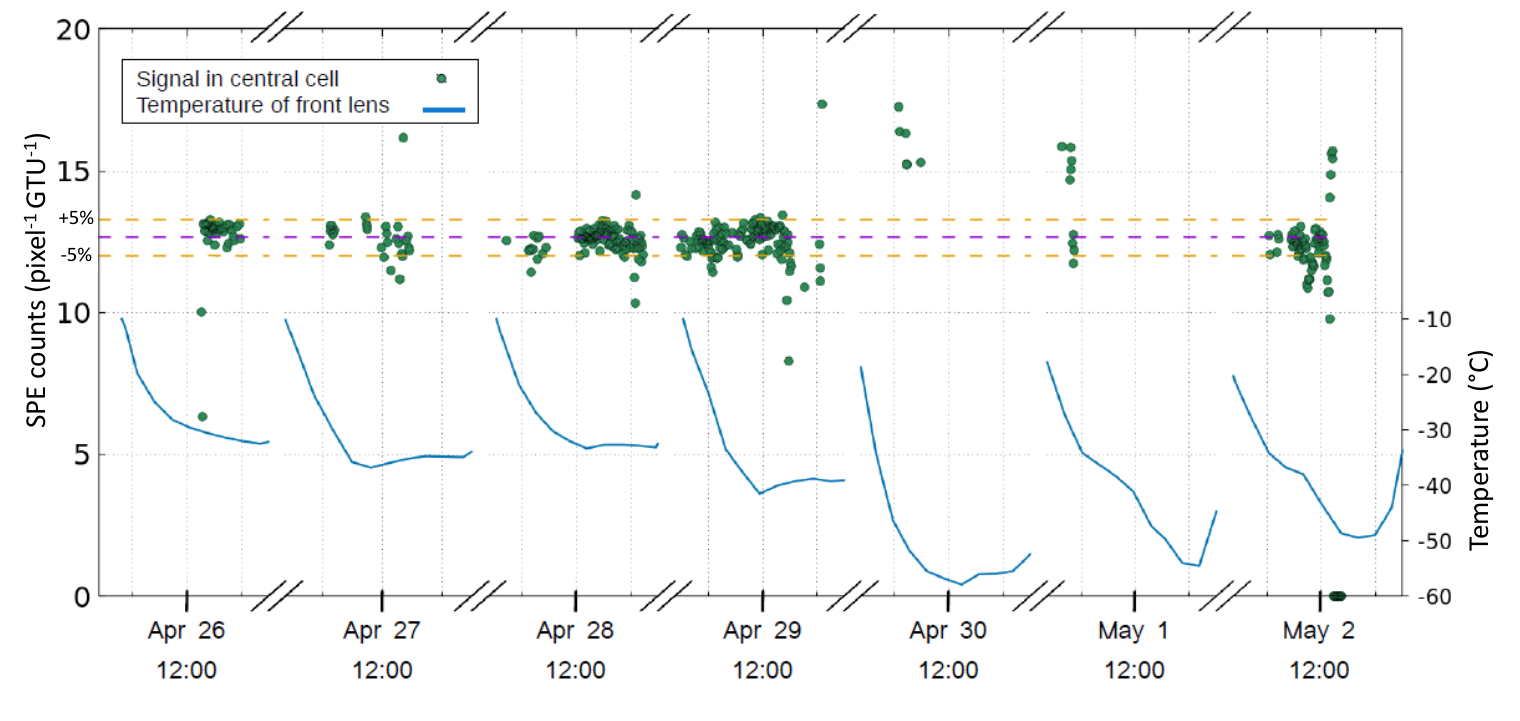} 
\caption{Response of the central cell of the EUSO-SPB1 PDM to flashes from the health LED (dots). The dashed lines indicate ${\pm}$5\% about the mean. The lower curves show the temperature measured at the outside of the Fresnel lens at the telescope optics aperture. Dates and times are in UT. Some gaps in the data occurred because the LED was turned off to avoid RF pickup from an onboard radio beacon that was enabled when the balloon descended below 21.3 km (70,000 ft).}
\label{fig:photo_stability}
\end{figure}

To monitor the PDM camera and readout function, a UV ``health" LED was fired twice every 16 s during flight (Figure~\ref{fig:Health_LED}). The sample of the LED measurements downloaded over the flight demonstrates the stability of the camera system response to the LED (Figure~\ref {fig:photo_stability}). Most of the data points fall within ${\pm 5\%}$ of the mean, despite nighttime temperature swings of 30${^\circ}$C outside the telescope.

\begin{figure} 
\begin{center}
\includegraphics [width=6.5in,keepaspectratio]{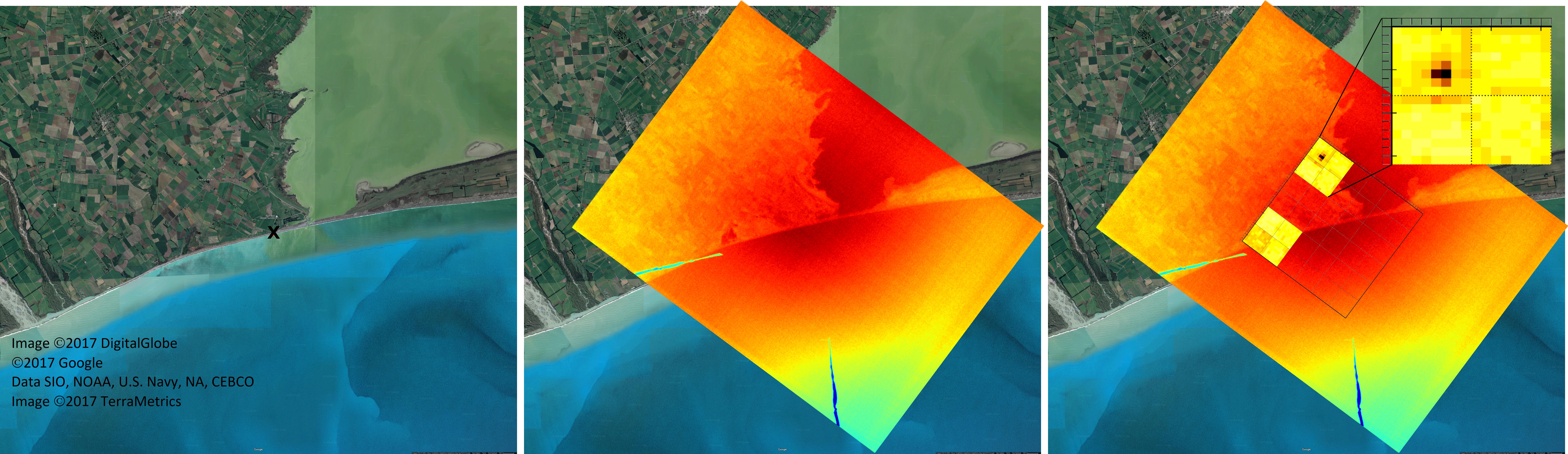}
\caption{At 8:32 UT, April 25, the balloon drifted over the New Zealand coast at an altitude of 33 km. Left: the X in the center of the satellite reference image \cite{2017retrieved..Ellis..Island} denotes the coordinates of the balloon: 43.86${^\circ}$S, 172.35${^\circ}$E. Center: the shape of the coast is reproduced in the overlaid image recorded by the IR camera. Its projected field of view on the ground as shown here is 14${\times}$17~km. Right: An unidentified ground light that the fluorescence telescope happened to observe (inset) provided additional in situ checks (see text).}
\label{fig:GSimage}
\end{center}
\end{figure}

The functionality of the IR camera and the telescope point-spread function were tested (through serendipity) as the balloon drifted over the eastern coast of the South Island of New Zealand about 8 hr after launch. An image from the IR camera shows the Pacific coastline and the edge of Lake Ellesmere in detail. The fluorescence telescope happened to be triggered by a light source on the ground (Figure~\ref{fig:GSimage}) and recorded a pixel pattern consistent with the expected point-spread function of the optics. To analyze the variability of the light source, we created a Lomb-Scargle \cite{1976ApSS..39..447L,1982ApJ...263..835S} periodogram consisting of 3000 consecutive measurements of the brightest pixel (after this time, the source moved to other pixels). The most prominent peak was at the period of 0.01~s. The corresponding 100 Hz frequency is consistent with the zero-volt crossing flicker of a fluorescent light bulb with an older magnetic ballast driven at 50 Hz, which is the frequency of the New Zealand electrical grid.https://www.overleaf.com/project/6487340c1579dcd88bac5c80
Using 100 Hz we were able to create a smooth phased light curve of the source. This analysis provided an in situ sanity check of the camera system's internal timing.

\begin{figure} 
\centering
\includegraphics[width=6.5in,keepaspectratio]{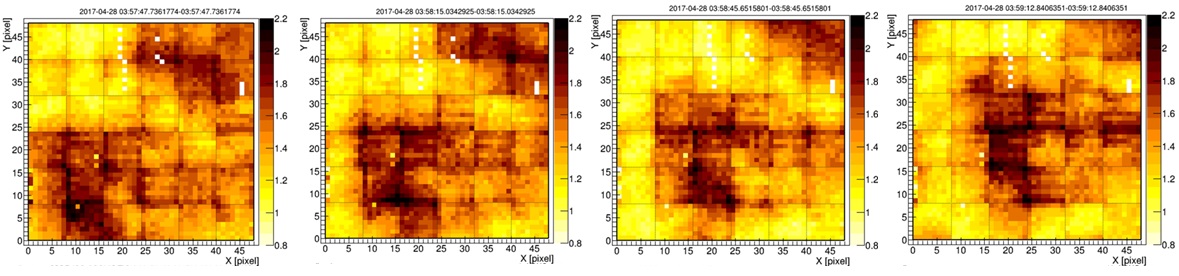} 
\caption{Sequence of time-averaged FT camera images that show clouds drifting across the field of view of the instrument. The four images in the figure are each separated by 30 s and each represents the average of about 1000~2.5~${\mu}$s images. Darker colors represent more SPE counts (light) recorded. The dark regions are clouds, and the observed average SPE counts from these regions are about 2~photoelectrons/pixel/2.5~${\mu}$s. The average count rates in the light-colored regions are about half this amount. These data were recorded on 2017 April 28.}
\label{fig:PDMclouds}
\end{figure}

Information about the presence of clouds in the PDM field of view can also be obtained from the PDM data. Variations in the average background rate can show clouds passing under the balloon, as demonstrated in Figure~\ref{fig:PDMclouds}. Clouds tend to be more reflective than the ground or ocean. At night they scatter light from airglow, stars and other sources, and appear in the FT as regions of higher average background rate.

 The telemetry bandwidth for data downloads was reduced when one of the two Iridium Pilot data links failed. Thirty of the 40 hr of data recorded on board were downloaded. The downloaded data included 175,000 recorded triggers. To optimize bandwidth prioritization, most of the data from the last 3 nights were not downloaded because the instrument was over high-cloud weather systems with poor viewing conditions. The flight was extended by controlled ballast releases. Unfortunately, the combination of the balloon leak and the emptied ballast hoppers led to an early controlled termination into the Pacific Ocean on May 6. Preparations had been underway to fly a Cessna 421C aircraft instrumented with UV LEDs and a UV laser under the balloon after one circumnavigation~\citep{mastafa:20173h}, but this didn't happen.

\section{Searching for Extensive Air Showers}

\begin{figure} [ht]
\centering
\includegraphics[width=5.0in,keepaspectratio]{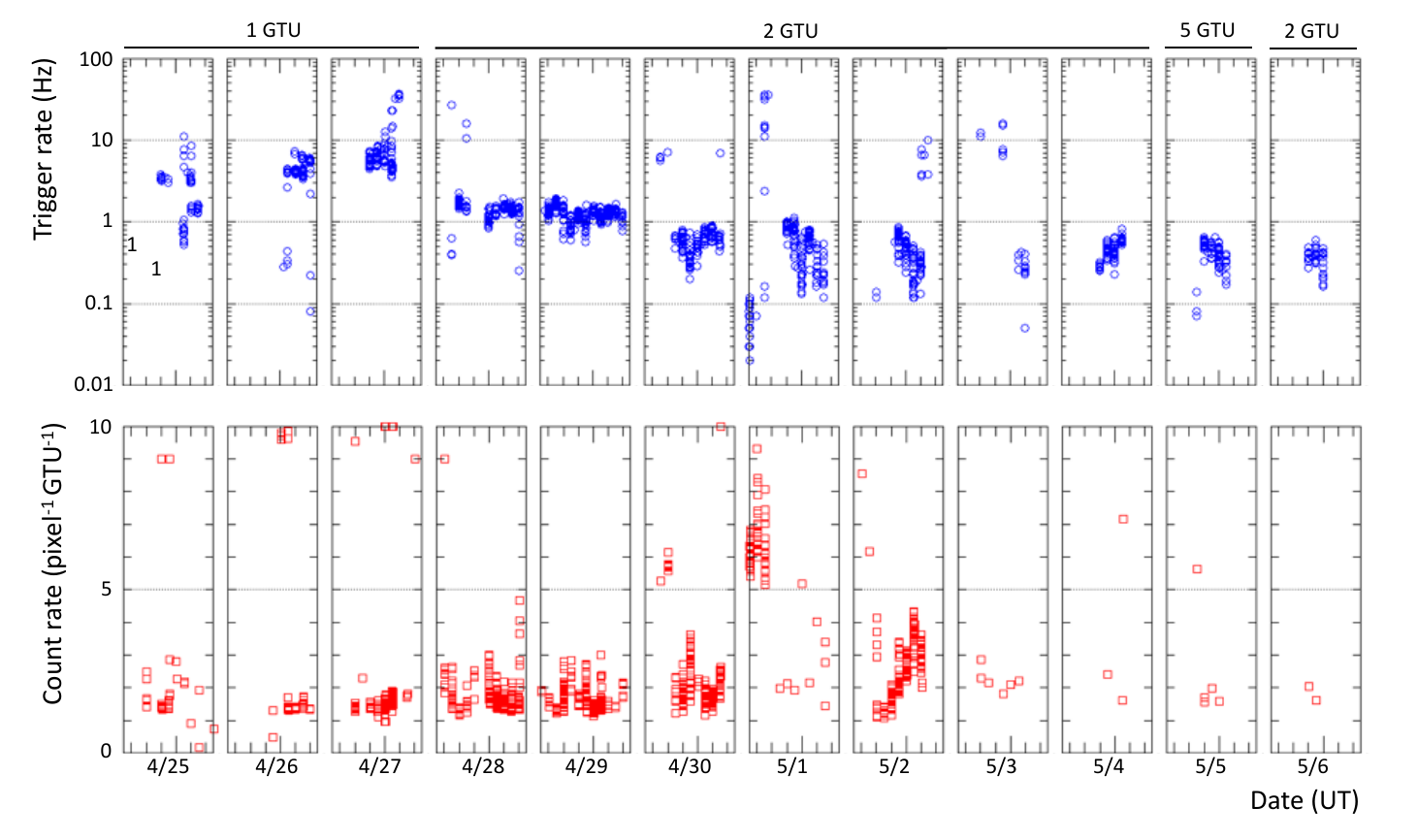} 
\caption{Top panel: The trigger rate measured during the mission. The trigger persistence setting is listed above. Bottom panel: The nighttime background light levels recorded during the mission. The units are SPE counts/pixel/2.5~${\mu}$s.}
\label{fig:trigger_bg_rates}
\end{figure}

The average background rates and trigger rates recorded during the mission are displayed in Figure~\ref{fig:trigger_bg_rates}. The higher trigger rates during the first 3 nights occurred while the PDM was operating with a 1~GTU persistence trigger setting. When the persistence setting was changed to 2~GTU, the typical trigger rate dropped below 2~Hz for the first 2 nights after the change, and then fell below 1~Hz for the rest of the mission.
 
\begin{figure} [htb]
\centering
\includegraphics[width=4.5 in,keepaspectratio]{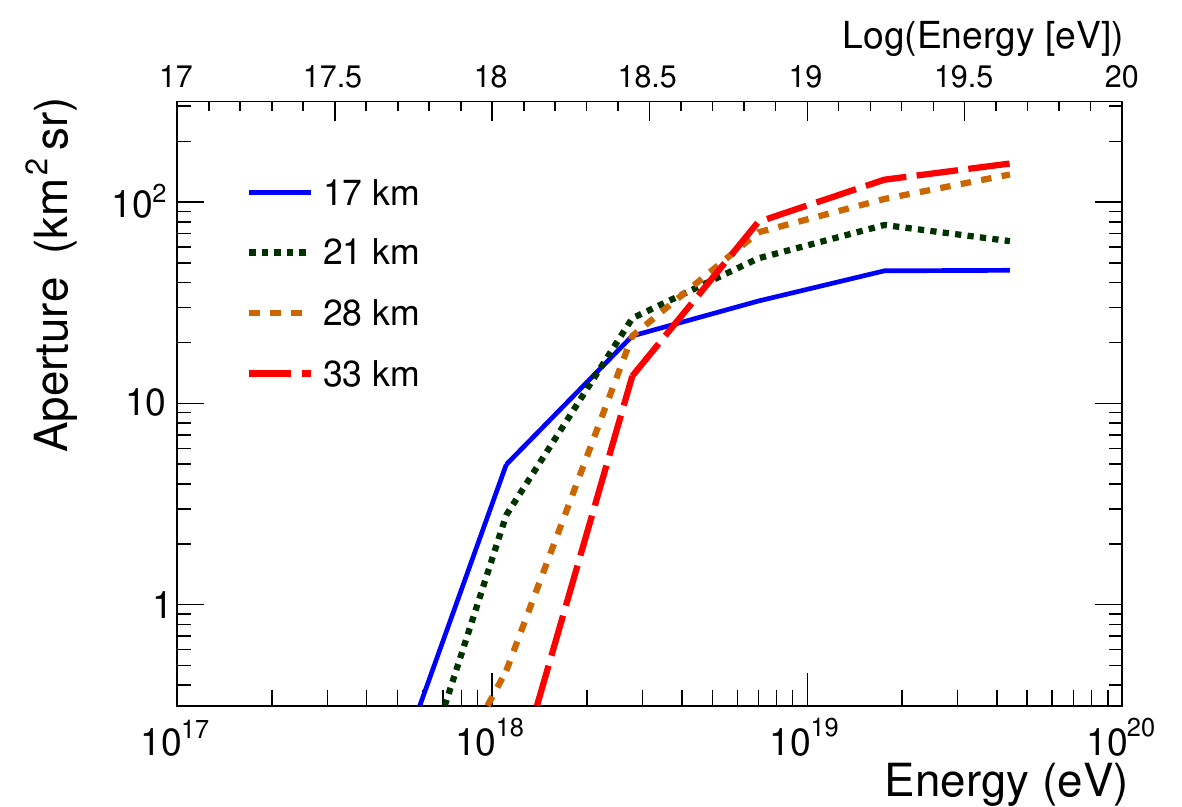} 
\caption{Estimated aperture of the EUSO-SPB1 fluorescence telescope as a function of primary particle energy (proton) for four altitudes that are representative of those encountered at night during data collection. The margin of statistical uncertainty of these estimates is $\le35\%$.
}
\label{fig:aperture}
\end{figure}

The instantaneous aperture was estimated as a function of air shower energy for four balloon altitudes (Figure \ref{fig:aperture}) that represent the periods when data were collected and analyzed. These altitudes ranged from the nominal 33 km float altitude, when the balloon envelope was in a super-pressure state, to 17 km, when the leaking balloon was descending over a cold storm system. The effect of the lower altitude was twofold. It reduced the highest energy aperture by about a factor of 3. It also lowered the energy threshold by about a factor of 2 because the telescope was closer to the troposphere, where nearly all EAS light production occurs.

\begin{figure} [htb]
\centering
\includegraphics[width=4.5 in,keepaspectratio]{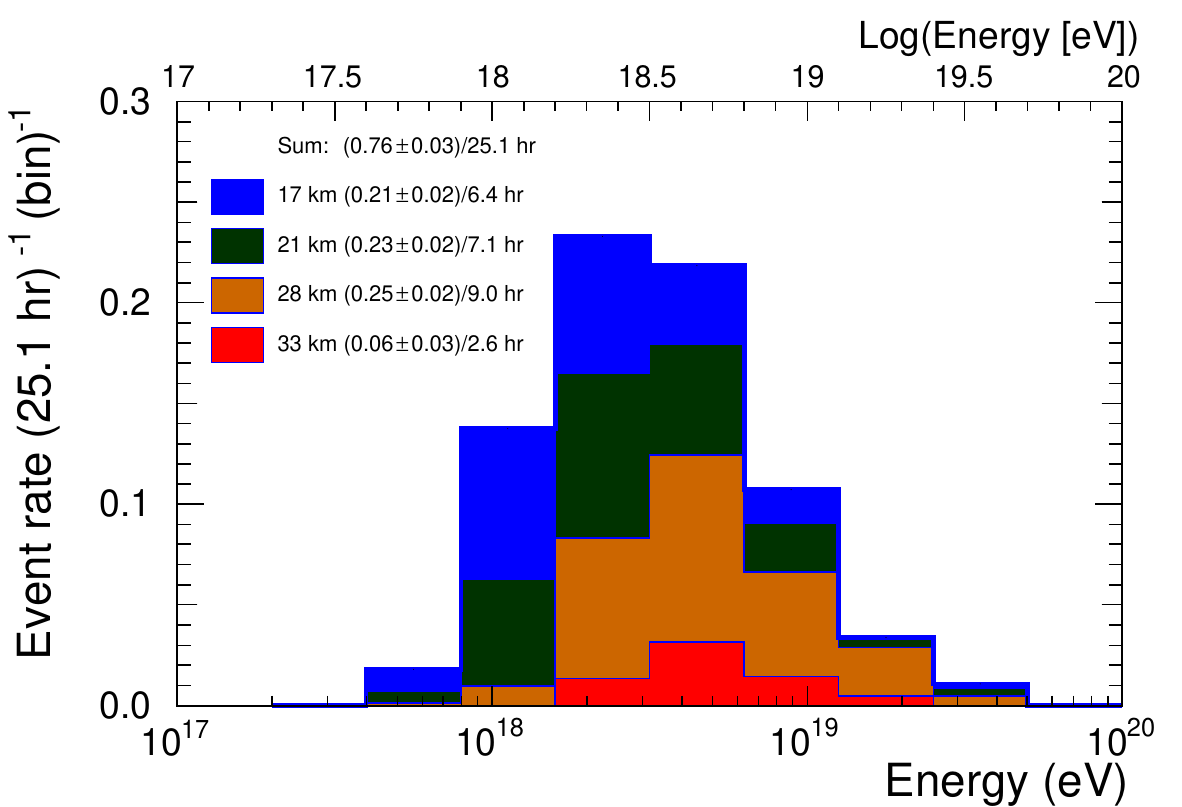}
\caption{Expected event rate distribution shown with the contributions for the four representative altitudes. This simulation assumed a clear atmosphere and low background conditions for an expected event rate of 0.76${\pm}$0.03 event over the 25.1 hr of data collected.
}
\label{fig:SimEASEnergy}
\end{figure}

A simulated energy distribution, based on the duration and nighttime altitudes of the SPB trajectory and the cosmic ray spectrum from~\citep{Fenu:2017OO}, is shown in Figure~\ref{fig:SimEASEnergy}. The distribution yields expected an event rate of 0.76${\pm}$0.03 event/25.1 hr when scaled to the duration of the mission flown, and assuming a clear atmosphere and low background conditions. (A total of 25.1 hr of FT data were downloaded and analyzed.) The uncertainty in the event rate is statistical, driven by the number of events in the simulation. For part of the mission, the balloon flew over high clouds that obscured the field of view. This effect~\citep{Bruno:2019IY,Monte:2019I4,Shinozaki:201955} was estimated to reduce the event rate by about a factor of 2 for an expected rate of 0.4 event/25 hr.

Two independent searches for EAS events were performed on this data set. One search~\citep{DiazDamian:2019Xe} involved scanning the data for triggers of potential interest, which were then classified into seven types. Of these triggers, 4128 were identified as having a duration between 3 and 50 GTUs and were visually scanned in detail. None showed a signature of a small or elongated cluster of pixels moving in a nearly straight line in the PDM at a speed consistent with that of an EAS, which moves through the atmosphere infinitesimally close to the speed of light. 
The second search for EAS events involved using a feature extraction method to form a simpler representation of an event, after which data were classified as EAS or noise using established machine learning techniques~\citep{Vrabel:2019hv}.
 A training data set combined simulated EAS samples and noise samples from EUSO-SPB1 data. The efficiency of the method was tested on laser tracks from the field campaign as a function of laser energy and on simulated EAS events as a function of primary particle energy. This search also did not yield any obvious EAS candidates.

\begin{figure} [htb]
\centering
\includegraphics[width=6.5in,keepaspectratio]{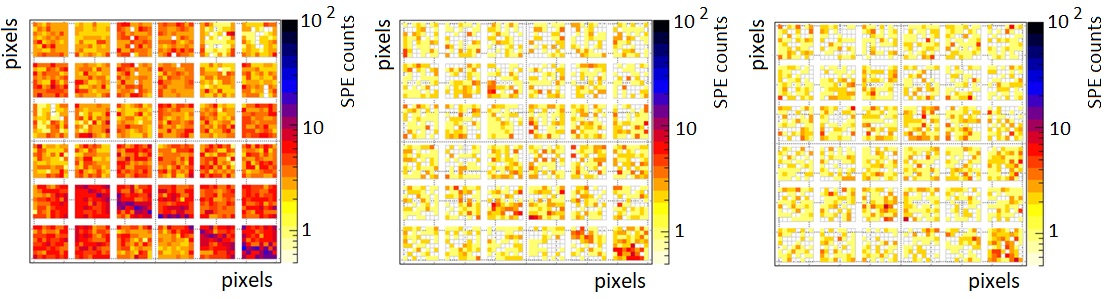} 
\caption{A direct cosmic ray traveling through the $48\times48$ pixel PDM camera. The three~panels correspond to three~consecutive 2.5~${\mu}$s time bins. The track structure in the first frame persists in two following time bins for this event. See text for discussion.}
\label{fig:directCR}.
\end{figure}

Both searches did identify background track-like triggers that appear to have been caused by very low energy cosmic particles, most likely muons, striking the PDM directly.  An example of one of these events is shown in the sequence of track-like images in Figure~\ref{fig:directCR}. A muon passing across the front face of the camera within in the optical filter and/or MaPMT windows could generate cherenkov light  Because these tracks cross the PDM within a single GTU time frame, they are readily distinguishable from an EAS track. An EAS track would require many GTUs in order to cross the PDM field of view because the fluorescence light is produced far below the telescope. In addition, an EAS track would not exhibit the persistence observed in this event (center and left panels of ~\ref{fig:directCR}),  The reason for this persistence is not well-understood. It may be due to a lingering ionization effect along the particle's path after it skimmed the front of the camera, for example. This persistence effect was not observed in the distant laser tracks recorded during field tests.

\section{EUSO-SPB2}
\label{sec:Future Missions}
Preparations are underway for a EUSO-SPB2 mission~\citep{Eser:2021H6} to expand the science goals beyond those of EUSO-SPB1 in support of a future space-based observatory for UHECRs and neutrinos, such as POEMMA.
EUSO-SPB2 will fly two astroparticle optical telescopes that use reflective optics~\citep{Kungel:20214E} to improve the optical efficiency relative to EUSO-SPB1. A fluorescence telescope~\citep{Osteria:2021OJ} with a field of view three times larger than that of EUSO-SPB1 and a 1~${\mu}$s GTU will point down to observe EAS tracks~\citep{Filippatos:2021a,Filippatos:2021b} and search for dark matter candidates ~\citep{Paul:2021W0}. A Cherenkov telescope~\citep{Bagheri:2021s0}, featuring a SiPM camera with 10~ns time bins, will point near Earth's limb to observe direct Cherenkov light from lower-energy cosmic rays above the limb~ \citep{PhysRevD.104.063029} and to search for neutrino signatures from tau neutrino interaction a few degrees below Earth's limb~\citep{PhysRevD.103.043017}. The optical backgrounds for such events are currently unexplored for suborbital altitudes. The balloon flight train will include an azimuth rotator. It will slew the gondola (and by extension the Cherenkov telescope) in azimuth to enable multi-messenger target-of-opportunity neutrino searches~\citep{PhysRevD.102.123013} in follow-up of selected international alerts from gravitational wave events, tidal disruption events, and gamma-ray bursts, for example. A 2023 launch from Wanaka is planned.

\section{Conclusions}
Although the EUSO-SPB1 mission of opportunity did not yield any cosmic-ray EAS events, most of the valid data were downloaded and analyzed. The data showed that the instrument performed well. The monitoring data from the health LED demonstrated the photometric stability of the camera at the ${\pm}$5\% level over the mission. The serendipitous observation of a ground light source on the first night of the mission demonstrated that the telescope, including the optics focusing, was operating as expected at float altitude. Because the payload was launched during the dark part of the moon cycle, the instrument searched for EASs every night of the flight. Although the balloon did not reach Argentina for a termination over land, the risk of test-flight anomalies was accepted to realize this target-of-opportunity mission. The null observation of EAS events was consistent with an expectation of about 0.4 EAS events for the data downloaded. This expectation value included an estimated factor-of-two reduction due to obscuration effects from high clouds. 

The mission raised the technical readiness level of the camera system flown and applied novel methods to test and characterize the fluorescence telescope in preflight field tests in the desert. Data from the field tests and the flight have also inspired the EUSO-SPB2 instrument design and mission planning.

\begin{acknowledgments}
This work was partially supported 
by NASA grants NNX13AH52G, NNX13AH53G, NNX13AH54G, NNX13AH55G, NNX16AG27G, 80NSSC18K0477, 80NSSC18K0464, 80NSSC18K0246, and 80NSSC19K0626;
the French Space Agency (CNES);
the Italian Space Agency (ASI) through the ASI INFN agreement 2017-8-H.0;
the Italian Ministry of Foreign Affairs and International Cooperation;
the Basic Science Interdisciplinary Research Projects of RIKEN and JSPS KAKENHI Grant (22340063, 23340081, and 
24244042);
the Deutsches Zentrum f\"ur Luft- und Raumfahrt and the Helmholtz Alliance for Astroparticle Physics (HAP) funded by the
Initiative and Networking Fund of the Helmholtz Association (Germany); 
National Science Centre Poland, grant 2017/27/B/ST9/02162.
This research used resources of the National Energy Research Scientific Computing Center (NERSC), 
a U.S. Department of Energy Office of Science user facility operated under contract no. DE-AC02-05CH11231.
We acknowledge the NASA Balloon Program Office and the Columbia Scientific Balloon Facility and their staffs for extensive support, and the Telescope Array Collaboration for the use of their facilities in Utah. We thank the Wanaka airport staff and manager Ralph Fegan. We also acknowledge the invaluable contributions of the administrative and technical staffs at our home institutions.

\end{acknowledgments}

\bibliography{EUSO-SPB1-AAS-APJ}{}
\bibliographystyle{aasjournal}

\allauthors
\end{document}